# The Convergence of European Business Cycles 1978 - 2000

**Paul Ormerod** (**Pormerod@volterra.co.uk**) *

and

**Craig Mounfield** (**Craig.Mounfield@volterra.co.uk**)

**Volterra Consulting Ltd**

**The Old Power Station**
**121 Mortlake High Street**
**London SW14 8SN**

**March 2001**

* Corresponding Author

*Abstract*

The degree of convergence of the business cycles of the economies of the European Union is a key policy issue. In particular, a substantial degree of convergence is needed if the European Central Bank is to be capable of setting a monetary policy which is appropriate to the stage of the cycle of the Euro zone economies.

We consider the annual rates of real GDP growth on a quarterly basis in the large core economies of the EU (France, Germany and Italy, plus the Netherlands) over the period 1978Q1 - 2000Q3. An important empirical question is the degree to which the correlations between these growth rates contain true information rather than noise. The technique of random matrix theory is able to answer this question, and has been recently applied successfully in the physics journals to financial markets data.

We find that the correlations between the growth rates of the core EU economies contain substantial amounts of true information, and exhibit considerable stability over time. Even in the late 1970s and early 1980s, these economies moved together closely over the course of the business cycle. There was a slight loosening at the time of German re-unification, but the economies are now, if anything, even more closely correlated.

As a benchmark for comparison, we add a series to the EU core data set which by construction is uncorrelated with these business cycles. We then analyse the EU core plus Spain, a country which has attached great importance to greater integration with Europe. In the early part of the period examined, the results are very similar to those obtained with the data set of the EU core plus the random series. However, there is a clear trend in the results, which provide strong evidence to support the view that the Spanish economy has now become closely converged with the core EU economies in terms of its movements over the business cycle.

In contrast, the results obtained with a data set of the EU core plus the UK show no such trend. In the late 1970s and early 1980s, the UK economy did exhibit some degree of correlation with those of the core EU. However, there is no clear evidence to suggest that the UK business cycle has moved more closely into line with that of the core EU economies over the 1978-2000 period.

1.   **Introduction**

Most of the countries of the European Union (EU) participated in the formation of the new currency, the Euro, on 1 January 1999.  Greece remained outside, but will shortly be joining, and Denmark rejected membership in a referendum in 2000.  The EU's second largest economy, the UK, remains outside and retains sterling as its currency.

A key feature of a monetary union such as the Euro is that monetary policy is common to all member states.  The structure of interest rates is effectively identical throughout the union. There may be small differences from state to state, but these are decidedly second-order.

It is therefore desirable that the economies of the member states of a monetary union should follow similar business cycles.  The level of interest rates appropriate in an economy which is experiencing a boom is unlikely to be so for an economy which is in recession.  At the time of writing, this issue has arisen with the Irish economy.  Ireland has been growing at exceptionally high rates, approaching 10 per cent a year, and as a result inflation in Ireland is above the target rate for inflation set by the European Central Bank.  But the level of interest rates in the Euro-zone is low and will not restrain growth in the Irish economy.

In practice, this is not a serious problem because the Irish economy is such a small part of the Euro-zone as a whole (around 1 per cent).  If, however, the business cycles of the major economies moved in different directions, serious policy problems would arise for the Central Bank.

This paper examines the extent to which the business cycles of the main EU states have been in synchronisation over the 1978 - 2000 period, and how this has altered over this period.  We examine the performance of the EU 'core', the large economies of France, Italy and Germany, which were founder members of both the EU itself and of the Euro and the core plus the large economy of Spain, which did not join the EU until 198x but which was a founder member of the Euro. This is contrasted with the core plus the UK, which whilst a member of the EU since 1973 has not joined the Euro and has been consistently the least supportive of ideas of further European integration.

We use the technique of random matrix theory (Mehta 1991) to analyse the correlations between the growth rates of the economies over time. Section 2 discusses the relevance of this theory, and section 3 sets out the empirical results.

## 2. Random matrix theory

Quarterly data exists for most of the EU economies over the past twenty years or so for the level of real output in the economy (GDP). We can therefore calculate annual growth rates quarter-by-quarter. The correlations between these growth rates for the various economies will inform us about the extent to which their business cycles are in synchronisation.

In other words, the degree of synchronisation of the business cycles may be quantified by calculation of the correlation matrix of the matrix of observations formed from the time series of GDP growth for each economy.

If $\underline{\underline{M}}$ is an N x T rectangular matrix (T observations of the GDP growth of the N economies) and $\underline{\underline{M}}^T$ is its transpose, the correlation matrix $\underline{\underline{C}}$ as defined below is an N x N square matrix

$$\underline{\underline{C}} = \frac{1}{T} \underline{\underline{M}}\, \underline{\underline{M}}^T$$

However due to the finite size of N (which corresponds to the number of economies) and T (which is the number of observations of GDP) then a reliable determination of the correlation matrix may prove to be problematic. The structure of the correlation matrix may be dominated by noise rather than by true information.

In order to assess the degree to which an empirical correlation matrix is noise dominated we can compare the eigenspectra properties of the empirical matrix with the theoretical eigenspectra properties of a random matrix. Undertaking this analysis will identify those eigenstates of the empirical matrix who contain genuine information content. The remaining eigenstates will be noise dominated and hence unstable over time. This technique has recently been applied by many researchers to financial market data (for example, Mantegna et al 1999, Laloux et al 1999, Plerou et al 1999, Gopikrishnan et al 2000, Plerou 2000, Bouchaud et al 2000, Drozdz et al 2001).

For a scaled random matrix **X** of dimension N x T, (i.e where all the elements of the matrix are drawn at random and then the matrix is scaled so that each column has mean zero and variance one), then the distribution of the eigenvalues of the correlation matrix of **X** is known in the limit T, N $\to \infty$ with Q = T/N $\geq$ 1 fixed (Sengupta et al 1999). The density of the eigenvalues of the correlation matrix, $\lambda$, is given by:

$$\rho(\lambda) = \frac{Q}{2\pi} \frac{\sqrt{(\lambda_{max} - \lambda)(\lambda - \lambda_{min})}}{\lambda} \qquad \text{for } \lambda \in [\lambda_{min}, \lambda_{max}]$$

and zero otherwise, where $\lambda_{max} = \sigma^2 (1 + 1/\sqrt{Q})^2$ and $\lambda_{min} = \sigma^2 (1 - 1/\sqrt{Q})^2$ (in this case $\sigma^2 = 1$ by construction).

The eigenvalue distribution of the correlation matrices of matrices of actual data can be compared to this distribution and thus, in theory, if the distribution of eigenvalues of an empirically formed matrix differs from the above distribution, then that matrix will not have random elements. In other words, there will be structure present in the correlation matrix.

To analyse the structure of eigenvectors lying outside of the noisy sub-space band the Inverse Participation Ratio (IPR) may be calculated. The IPR is commonly utilised in localisation theory to quantify the contribution of the different components of an eigenvector to the magnitude of that eigenvector (thus determining if an eigenstate is localised or extended) (Plerou et al 1999).

Component $i$ of an eigenvector $v_i^\alpha$ corresponds to the contribution of time series $i$ to that eigenvector. That is to say, in this context, it corresponds to the contribution of economy $i$ to eigenvector $\alpha$. In order to quantify this we define the IPR for eigenvector $\alpha$ to be

$$I^\alpha = \sum_{i=1}^{N} (v_i^\alpha)^4$$

Hence an eigenvector with identical components $v_i^\alpha = 1/\sqrt{N}$ will have $I^\alpha = 1/N$ and an eigenvector with one non-zero component will have $I^\alpha = 1$. Therefore the inverse participation ratio is the reciprocal of the number of eigenvector components significantly different from zero (i.e. the number of economies contributing to that eigenvector).

### 3. The data and the results

Quarterly levels of real GDP over the period 1977Q1 - 2000Q3 are available from the OECD database for the largest EU economies, France, Germany, Italy, Spain and the UK. The first three plus the Benelux countries are widely regarded as forming the EU 'core', being the founder members of the (then) European Economic Community. Quarterly data is available for the Netherlands but not for Belgium, and we include the former in the 'core' group[1].

We analyse the correlation matrix of real GDP growth rates for the following groups of countries:

- EU 'core' i.e. France, Germany, Italy and the Netherlands
- EU core plus Spain

- EU core plus the UK

As a comparator, we also analyse the EU core plus a random data series generated from a large number of random shuffles of the data for Germany. This sets the benchmark of what we would expect to see if an economy were added to the EU core data set whose short-term growth rates over the business cycle are by construction not correlated with those of the core members.

In terms of the EU core, there is a large amount of genuine correlation between the growth rates of the economies over the business cycle. Further, there is a substantial degree of stability of these correlations over the 1978-2000 period.

The theoretical range of the eigenvalues for a random matrix of the relevant order is between 0.62 and 1.46. The eigenvalues of the empirical correlation matrix of annual growth rates over the 1978Q1 - 2000Q3 period are 2.68, 0.69, 0.39 and 0.24, indicating the presence of a large amount of true information in the correlation matrix.

An illustration of the stability of the correlation matrix is given by the following. For those eigenvectors that deviate from the theoretically predicted bounds of random matrix theory it is important to quantify the degree of stability of the information content of the eigenmode. We may assess this stability by calculating the scalar product of eigenvectors in non-overlapping analysis periods. That is for two non-overlapping series of observations of GDP growth $T_A$ and $T_B$ we form the overlap matrix

$$\underline{\underline{O}}(T_A, T_B) = \begin{pmatrix} \vec{v}^N(T_A) \cdot \vec{v}^N(T_B) & . & . & . & \vec{v}^N(T_A) \cdot \vec{v}^1(T_B) \\ & . & . & . \\ & . & . & . \\ & . & . & . \\ \vec{v}^1(T_A) \cdot \vec{v}^N(T_B) & . & . & . & \vec{v}^1(T_A) \cdot \vec{v}^1(T_B) \end{pmatrix}$$

---

[1] The Luxemburg economy is trivially small

Hence if the eigenvector structure remains perfectly stable in time (i.e. the correlations between the assets contributing to that eigenvector remain stable from period to period) then each element of the overlap matrix would be equal to $O_{ij}(T_A,T_B) = \delta_{ij}$. No inter-period stability would imply that $O_{ij}(T_A,T_B) = 0$.

Figure 1(a) shows the overlap matrix for the EU core countries, with the data split into two (almost) equal periods from 1978Q1 - 1988Q4 and from 1989Q1 - 2000Q3.

The plot is colour coded so that black corresponds to an overlap of 0 and white corresponds to an overlap of 1. The axes are arranged so that the overlap of eigenvector 1 (corresponding to the largest eigenvalue) between the two periods is in the bottom right hand corner of the plot (i.e. the same as that in the matrix given above), and the overlap of eigenvector N (the smallest eigenvector) between the two periods is in the top left hand corner.

Figure 1(a) demonstrates that the largest eigenvector displays a high degree of stability between the two periods (the observed overlap was 0.99) but that the smallest eigenvectors display less stability. For comparison, in figure 1(b) is plotted the same overlap matrix for the same 4 time series, but in this case the 4 time series have been shuffled at random 100,000 times. This has the effect of destroying any temporal correlations present in the structure of the data.

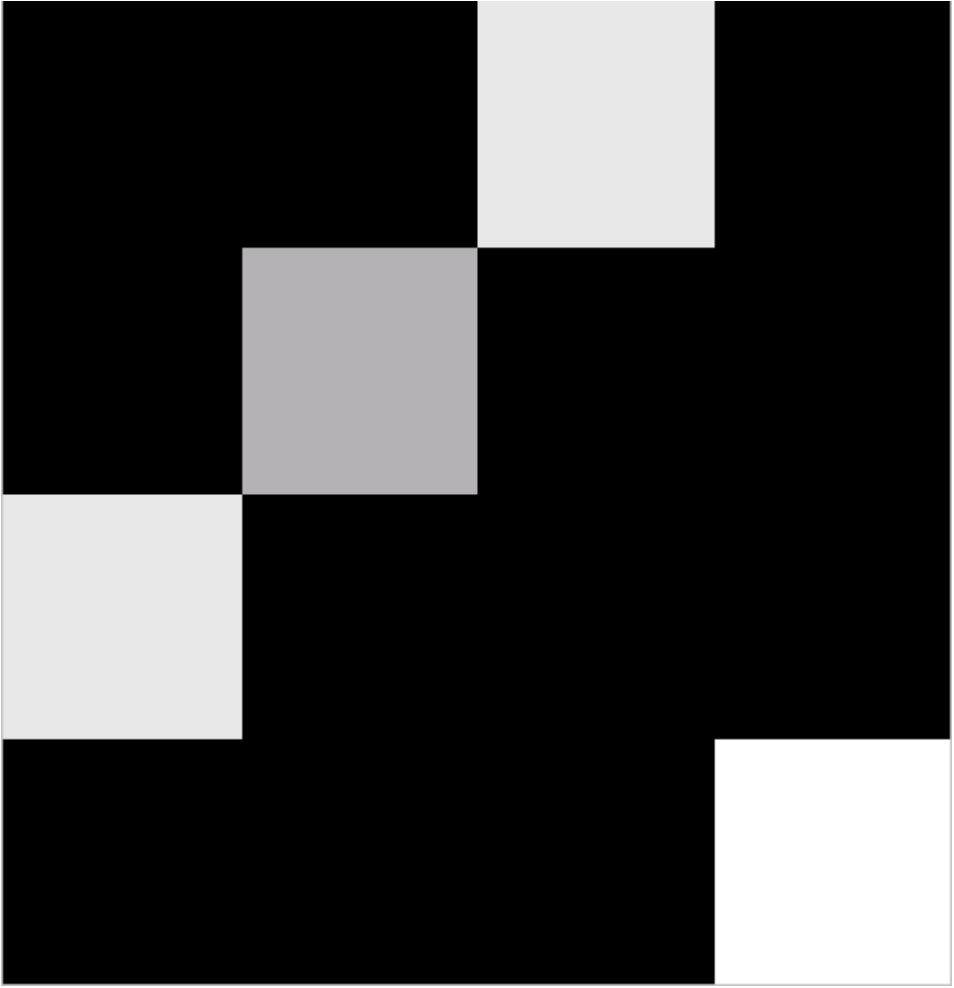

**Figure 1(a)**

*Colour coded plot of the degree of overlap of the eigenvectors corresponding to 2 non-overlapping analysis periods for the core EU economies of France, Germany, Italy and the Netherlands. A white square corresponds to perfect overlap between the structure of the 2 eigenvectors (perfect stability of the degree of information content in that eigenmode) and black corresponds to no degree of overlap whatsoever. As can be seen, the degree of stability of the market eigenmode (i.e. the dot product of eigenvector 1 with itself in each of the two periods - bottom right hand corner) is significantly different from that of any of the other overlaps.*

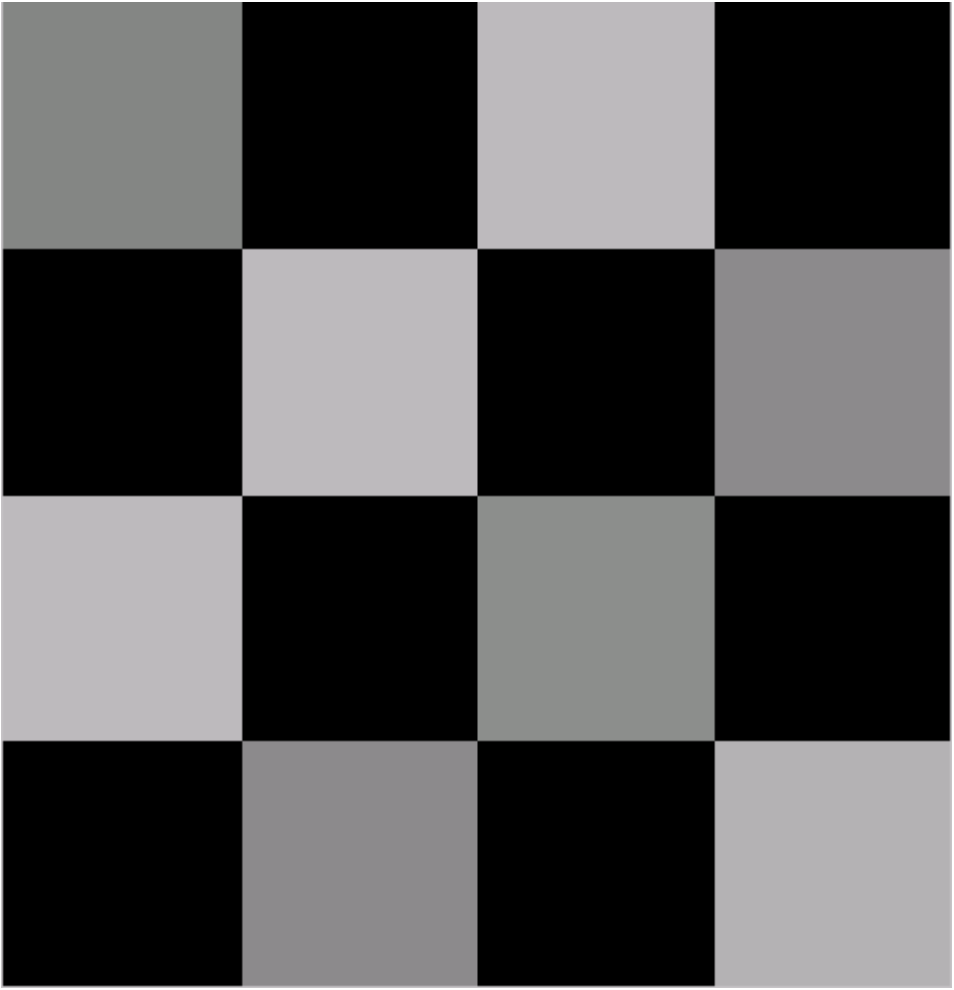

**Figure 1(b)**

*Colour coded plot of the degree of overlap of the eigenvectors corresponding to 2 non-overlapping analysis periods for the core EU economies of France, Germany, Italy and the Netherlands. In this case the 4 time series have been shuffled at random 100,000 times destroying any temporal correlation in changes in GDP growth. As can be seen the distribution of stability of information content is significantly different from that in figure 1(a). This demonstrates that the temporal structure of the information contained within the correlation matrix is stable over long periods of time*

In terms of those eigenvalues which lie outside the noisy sub-space band the most important from a macroeconomic perspective is the largest eigenvalue. The application of these techniques to equities traded in financial markets have demonstrated that this eigenmode corresponds to the 'market' eigenmode (e.g. Gopikrishnan et al, 2000). In this context the largest eigenvalue will inform us as to the degree to which the movements of the EU economies are correlated.

The contribution which each of the core economies makes to eigenvector 1 can be seen from calculating the IPR. The components are in fact (0.49, 0.55, 0.51, 0.44), which gives a calculated value of the IPR of 0.256, indicating that all four economies are contributing approximately equally to this eigenvector.

The fact that the individual elements are almost identical in size also shows that this vector corresponds to a collective motion of all of the GDP growth time series. It is therefore a measure of the degree to which the growth of different countries in the EU core is correlated.

The trace of the correlation matrix is conserved, and is equal to the number of independent variables for which time series are analysed. That is, for the core EU correlation matrix the trace is equal to 4 (since there are 4 time series). The closer the 'market' eigenmode (i.e. eigenmode 1) is to this value the more information is contained within this mode i.e. the more correlated the movements of GDP. The market eigenmode corresponds to the largest eigenvalue. The degree of information contained within this eigenmode, expressed as a percentage, is therefore $100\lambda_{max}/N$.

To follow the evolution of the degree of business cycle convergence over time we may analyse how this quantity evolves temporally. The analysis is undertaken with a fixed window of data. Within this window the spectral properties of the correlation matrix formed from this data set are calculated. In particular the maximum eigenvalue is calculated. This window is then advanced by one period and the maximum eigenvalue noted for each period.

The choice of an appropriate window to span the periodicity of what constitutes the business cycle is not completely straightforward. Business activity is influenced by a very large number of events, and these events may be very diverse in character and scope.

Individual cycles therefore vary both in terms of amplitude and period. This lack of regularity may be analysed formally using random matrix techniques (Ormerod and Mounfield 2000). The evidence for the existence of a business cycle at all relies more upon factors such as the fact that output changes in different sectors of an economy tend to move together (Lucas 1977) than upon regularities in either amplitude or period of the economy as a whole.

A major study of the US economy (Burns and Mitchell 1946) many years ago concluded that the period ranged from some two to twelve years, a range which still commands broad assent amongst economists, though the upper bound might now be felt to be slightly high. We initially carried out results for a window of 10 years, although the results for a window of 8 years are virtually identical, and it is these which we present here. The results are in fact robust to the choice of window, until a window as short as 5 years is chosen, when greater instability begins to be introduced, due to measurement noise induced by the reduced number of observations.

The results for the core EU economies are set out in Figure 2. Each window contains 32 quarterly observations, and so we have 60 windows in total. The period 1978Q1 - 1985Q4 corresponds to the first data point in Figure 2, 1978Q2 - 1986Q1 to the second, and so on through to 1992Q4 - 2000Q3.

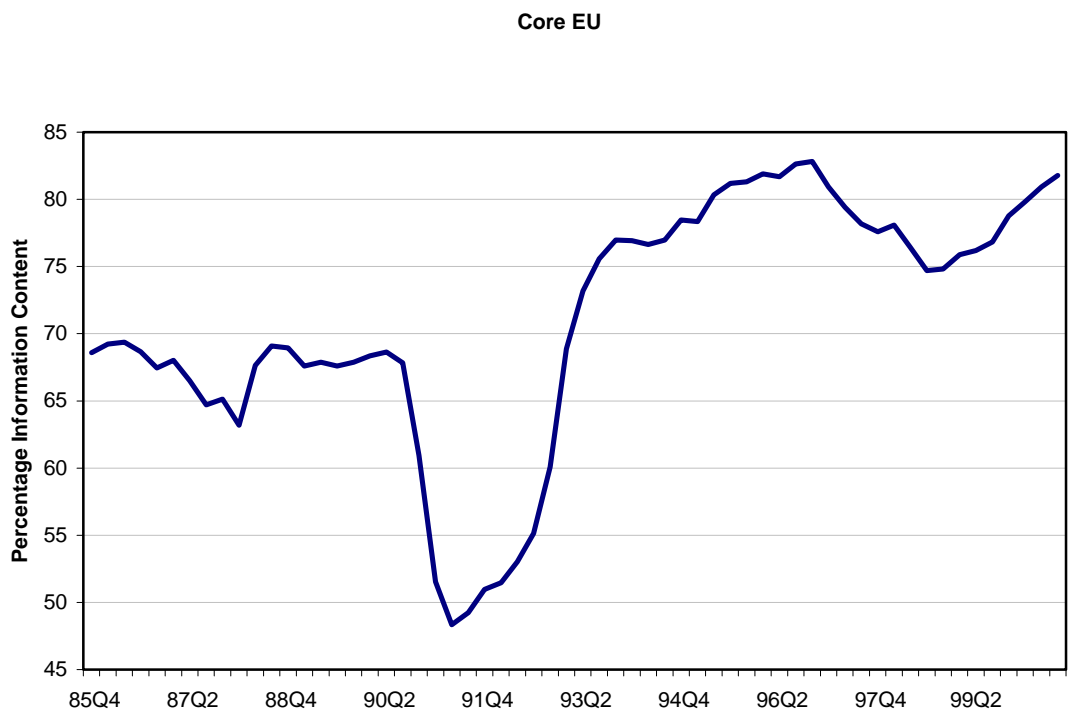

**Figure 2**

*The temporal evolution of the degree of information content in the maximum eigenvalue of the empirical correlation matrix formed from the time series of quarterly GDP growth for the core EU economies of France, Germany, Italy and the Netherlands.*

Even in the early part of the period, the 'market' eigenvalue took up some 70 per cent of the total of the eigenvalues, indicating a strong degree of convergence of the business cycles of the EU core economies. There was a temporary reduction of convergence around the time of German re-unification in the early 1990s, but the economies rapidly re-converged and the principal eigenvalue now accounts for some 80 per cent of the total information content within the correlation matrix, indicating a movement towards even greater convergence of the business cycles of the EU core economies over time.

As a benchmark for comparison, Figure 3 illustrates the effect of adding a purely random series to the core EU data set. This consists of the German quarterly growth rate data shuffled 100,000 times (thereby destroying any temporal correlations in the data).

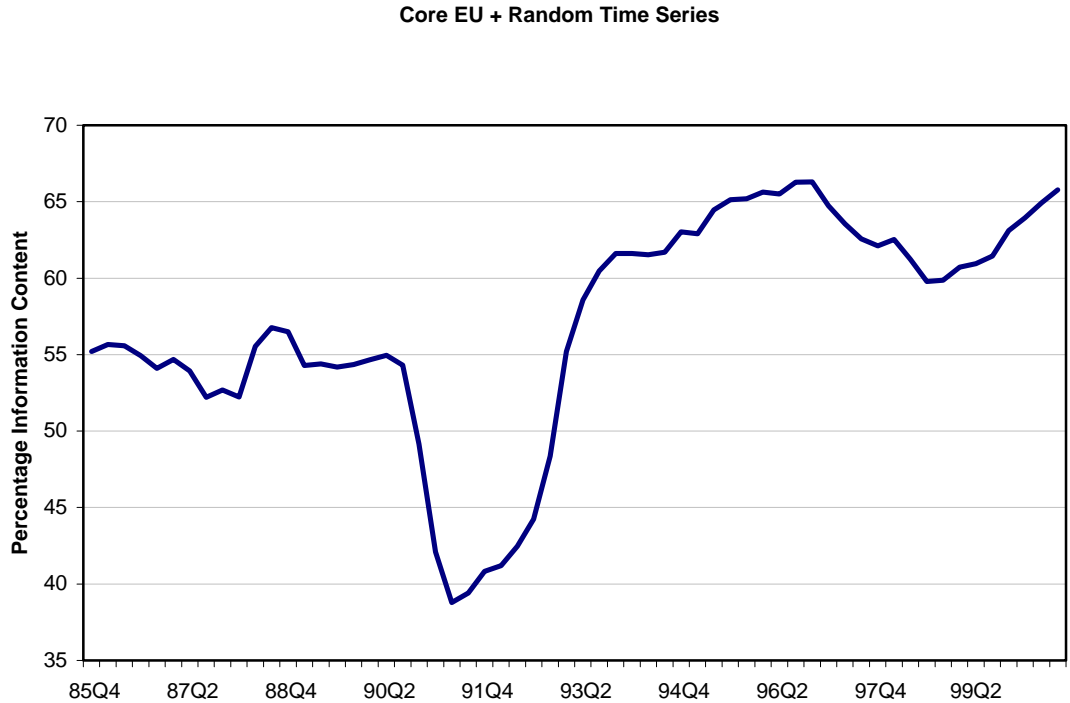

**Core EU + Random Time Series**

**Figure 3**

*The temporal evolution of the degree of information content in the maximum eigenvalue of the empirical correlation matrix formed from the time series of quarterly GDP growth for the core EU economies of France, Germany, Italy and the Netherlands plus a time series formed from 100,000 random shuffles of the German GDP time series*
*.*

The pattern of movement is almost identical to that of Figure 2. The important point to note from this is the scale over which the contribution of the maximum eigenvalue moves. There are now 5 series in the data set rather than 4, so the sum of the eigenvalues is now 5. Essentially, the data plotted in Figure 3 is the data in Figure 2 multiplied by 4/5.

In other words, Figure 3 represents what we would observe if an economy whose business cycle were completely uncorrelated to that of the EU core were added to the data set.

We now move on to examine the case of Spain. After many years isolated under dictatorship, the Spanish authorities have attached great importance to modernising their economy and society in a European context. Policy has been strongly supportive of European integration. The extent to which business cycle convergence has been achieved with the EU core is plotted in Figure 4.

**Core EU + Spain**

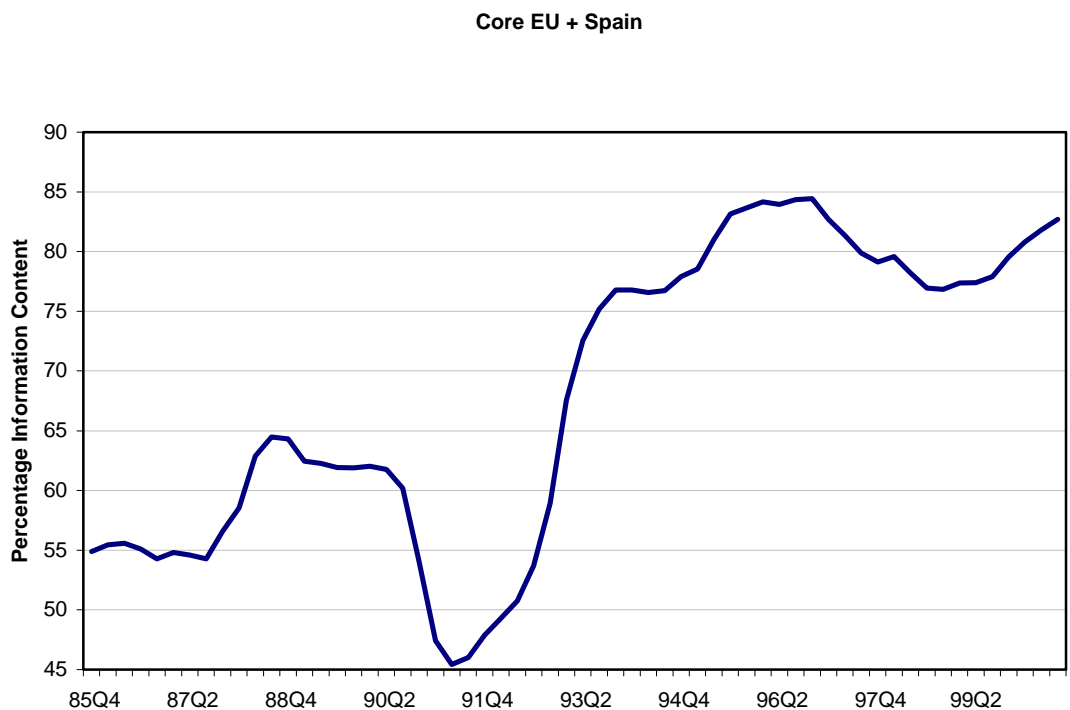

**Figure 4**

*The temporal evolution of the degree of information content in the maximum eigenvalue of the empirical correlation matrix formed from the time series of quarterly GDP growth for the core EU economies of France, Germany, Italy and the Netherlands plus the time series of GDP growth for the Spanish economy.*

Qualitatively, the pattern over time is similar to that of Figure 2, reflecting, for example, the temporary impact of German re-unification. But there is a very clear upward trend in these results. In the early parts of the window, the value of $100\lambda_{max}/N$ is around 55, very similar to that of the core EU plus a random data series. However, by the end this has risen to around 80, very similar to that of the core EU alone.

In other words, this suggests strong evidence to support the view that the Spanish economy has become closely converged with the core EU economies in terms of its movements over the business cycle.

In contrast, Figure 5 shows the results for the core EU plus the UK.

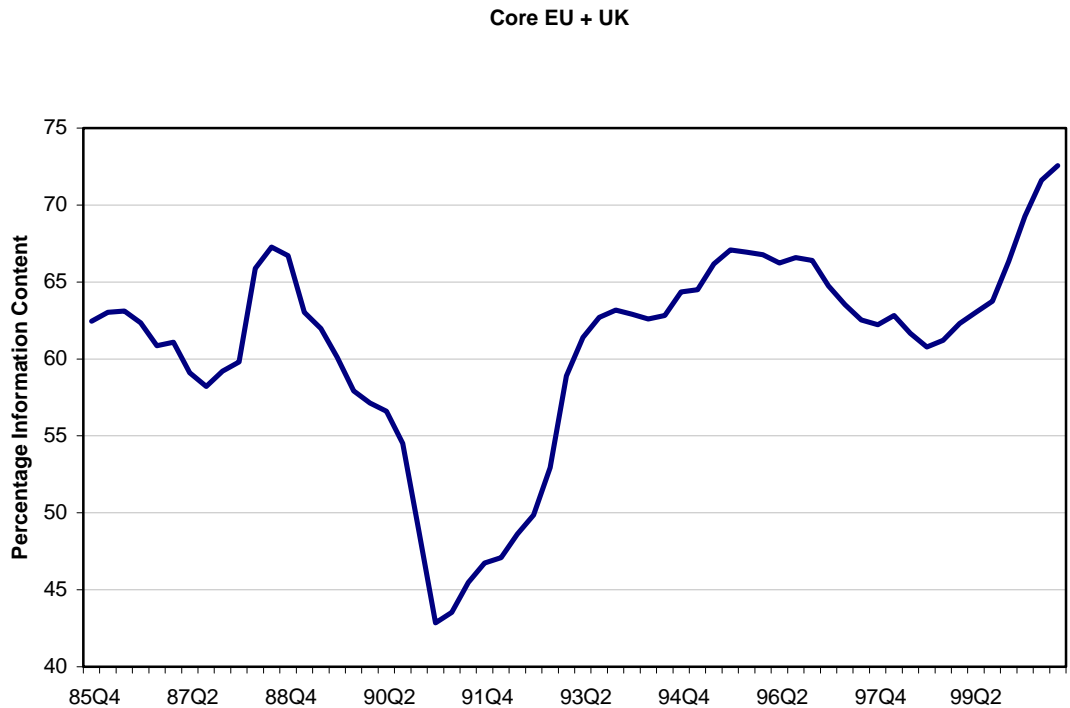

**Figure 5**

*The temporal evolution of the degree of information content in the maximum eigenvalue of the empirical correlation matrix formed from the time series of quarterly GDP growth for the core EU economies of France, Germany, Italy and the Netherlands plus the time series of GDP growth for the UK economy.*

In the early part of the analysis period, the value of $100\lambda_{max}/N$ for the EU core plus UK is around 65, less than for the core EU itself, but distinctly higher than for either the core EU plus Spain or the core EU plus a random series. However, subsequently the value shows no clear trend, unlike the case when Spain is added. At the very end of the analysis period, there is a rise to just over 70, but this remains well below the value for the EU core and the EU core and Spain, and indeed may simply be a temporary fluctuation around an average value of some 65.

In short, there is certainly no clear evidence to suggest that the UK business cycle has moved more closely into line with that of the core EU economies over the 1978-2000 period.

Table 1 summarises these findings

|  | **Average Information Content in Market Eigenmode in First 20 Periods** | **Average Information Content in Market Eigenmode in Last 20 Periods** |
|---|---|---|
| **Core EU** | 68% | 79% |
| **Core EU + Random** | 55% | 63% |
| **Core EU + Spain** | 59% | 81% |
| **Core EU + UK** | 61% | 65% |

**Table 1**

4. **Conclusion**

In this paper, we analyse the convergence or otherwise of the business cycle in the main economies of the European Union, using the annual growth rates of quarterly real GDP over the 1978Q1 - 2000Q3 period. The correlations between the growth rates are analysed using random matrix theory, which enables us to identify the extent to which the correlations contain true information rather than noise.

For the core EU countries, France, Germany, Italy and the Netherlands, we find that the business cycles have shown strong synchronisation over the whole of the 1978-2000 period. Further, the correlations between them are stable over time, though if anything they have become stronger during the 1990s.

As a benchmark for comparison, we analyse the core EU data plus a series which by construction is not correlated with this data. We then examine the core EU plus Spain. In the early part of the period, the results are very similar to those obtained when the uncorrelated data is added. But there is a clear trend in the results which supports the view that the Spanish economy has become closely converged with the core EU economies in terms of its movements over the business cycle.

In contrast, the results obtained when the UK is added to the core EU data set exhibit no such trend, and there is certainly no clear evidence to suggest that the UK business cycle has moved more closely into line with that of the core EU economies over the 1978-2000 period.